\documentclass[aip,cha,reprint,numerical,onecolumn]{revtex4-1}
\usepackage{graphicx}
\usepackage{amsmath}
\usepackage{amssymb}
\usepackage[hidelinks]{hyperref}

\def\re{\mbox{Re}}
\def\im{\mbox{Im}}

\def\w{\omega}
\def\W{\Omega}
\def\t{\tau}
\def\e{\varepsilon}
\def\vp{\varphi}
\def\Vp{\Phi}
\def\r{\rho}
\def\T{\Theta}
\def\t{\theta}
\def\p{\psi}

\begin{document}
\title{Solitary States and Partial Synchrony in Oscillatory Ensembles with
 Attractive and Repulsive Interactions}
\date{\today}
\author{Erik Teichmann}
\email{kontakt.teichmann@gmail.com}
\affiliation{Institute of Physics and Astronomy, University of Potsdam,
Karl-Liebknecht-Str. 24/25, 14476 Potsdam-Golm, Germany}
\author{Michael Rosenblum}
\affiliation{Institute of Physics and Astronomy, University of Potsdam,
Karl-Liebknecht-Str. 24/25, 14476 Potsdam-Golm, Germany}
\affiliation{Control Theory Department, Institute of Information Technologies,
Mathematics and Mechanics, Lobachevsky University Nizhny Novgorod, Russia}

\date{\today}

\begin{abstract}
We numerically and analytically analyze transitions between different
synchronous states in a network of globally coupled phase oscillators with
attractive and repulsive interactions. The elements within the attractive or
repulsive group are identical, but natural frequencies of the groups differ.
In addition to a synchronous two-cluster state, the system exhibits a solitary
state, when a single oscillator leaves the cluster of repulsive elements, as
well as partially synchronous quasiperiodic dynamics. We demonstrate how the
transitions between these states occur when the repulsion starts to prevail
over attraction.
\end{abstract}


\maketitle

\begin{quotation}
Networks of coupled oscillators are a popular model for many engineered
or natural systems.
The main effect -- emergence of a collective mode via synchronization -- is now well-understood
and therefore focus of research shifted recently to analysis of different complex states.
These states
include chimeras, when a population of identical units splits into a
synchronous and asynchronous part, quasiperiodic partially synchronous states,
characterized by the difference of frequencies of individual units and of the
collective mode, and clusters and heteroclinic cycles, to name
just a few.  Of particular interest are ensembles where some elements have
only attractive connections while others have only repulsive ones. This model
is motivated by studies of neuronal networks that are built from excitatory
and inhibitory neurons. In this paper we analyze how the state of such a setup
changes with the interplay of attraction and repulsion.  We demonstrate that
if the frequency mismatch between attractive and repulsive units is smaller than
some critical value then desynchronization occurs via appearance of the solitary state.
With the further increase of repulsion the system undergoes a transition to
quasiperiodic  partial synchrony.  In the latter state the attractive units
remain synchronized, while the repulsive group settles between synchrony and
asynchrony so that the mean fields of both groups remain locked, but the frequency of the
repulsive elements is larger than that of their mean field.
For a large frequency mismatch of attractive and repulsive groups desynchronization
immediately leads to partial synchrony.
\end{quotation}

\section{Introduction}
Investigation of coordinated dynamics of many interactive oscillatory elements
is relevant for the understanding of various phenomena from different branches of
science.  Probably, the most important and also mostly studied effect is the
emergence of a collective mode, observed in populations of flashing
fireflies~\cite{kaempfer1906history}, groups of pedestrians on
footbridges~\cite{strogatz2005theoretical} or metronomes placed on a common
support~\cite{martens2013chimera}, electronic
circuits~\cite{watanabe1994constants}, populations of
cells~\cite{richard1996acetaldehyde}, synthetic genetic
oscillators~\cite{prindle2011sensing}, etc.  Besides of collective synchrony,
oscillatory networks exhibit many other interesting dynamical states like
clusters and heteroclinic switching~\cite{hansel1993clustering},
chimeras~\cite{kuramoto2002coexistence}, collective
chaos~\cite{hakim1992dynamics}, traveling waves~\cite{hooper1988travelling},
quasiperiodic partial synchrony~\cite{vanvreeswijk1996partial,
rosenblum2007self, pikovsky2009self, clusella2016minimal}, solitary
states~\cite{maistrenko2014solitary}, and so on.  Analysis of such states and
transitions between them is in the focus of current research.

Some of mentioned effects can be studied within the framework of the famous
Kuramoto model~\cite{kuramoto1984chemical} and of its immediate extension, the
Kuramoto-Sakaguchi model~\cite{sakaguchi1986soluble}, that treat phase
oscillators with the sine-coupling.  Though this is a rather simplistic
description of real-world oscillators, these models became extremely popular due
to the possibility of analytical treatment~\cite{acebron2005kuramoto,
pikovsky2015dynamics}.  For example, they allow for theoretical description of
synchronization transitions (that, in dependence on the distribution of
oscillatory frequencies, can be alike second- or first-order
\cite{pazo2005thermodynamic} phase transitions).  Due to their specific
mathematical properties, sine-coupled phase oscillators also often admit a
low-dimensional description via the Watanabe-Strogatz
(WS)~\cite{watanabe1993integrability, watanabe1994constants} and Ott-Antonsen
(OA)~\cite{ott2008low, ott2009long} theories. All this explains why the
Kuramoto-Sakaguchi model became a paradigmatic one, with applications ranging
from explanation of social effects~\cite{kaempfer1906history,
strogatz2005theoretical} to neuroscience~\cite{breakspear2010generative}.

In most variants of the Kuramoto-Sakaguchi model researchers treat networks with
attractive interactions and the existing literature extensively covers this
case~\cite{montbrio2004synchronization, abrams2008solvable,
barreto2008synchronization}.  Networks of repulsive elements attract much less
attention, although they show interesting
effects~\cite{vanvreeswijk1994inhibition, tsimring2005repulsive,
pimenova2016interplay}.  Not much attention is also paid to mixed
networks~\cite{hong2011kuramoto, hong2011conformists, anderson2012multiscale,
iatsenko2013stationary, vlasov2014synchronization, qiu2016synchronization},
consisting of both attractive and repulsive elements, though systems of this
type are common in neuroscience, because real neurons interact via excitatory
and inhibitory connections~\cite{wilson1972excitatory, vreeswijk1996chaos,
peyrache2012spatiotemporal, dehghani2016dynamic}.

In this paper we concentrate on emergence of solitary state and quasiperiodic
partial synchrony in networks with attractive and repulsive connections.  The
solitary state, when a single repulsive unit leaves the synchronous cluster, was
for the first time found and analyzed  in  Ref.~\cite{maistrenko2014solitary}
and later in Refs.~\cite{brezetskyi2015rare, jaros2015chimera,
chouzouris2018chimera, chen2019dynamics, majhi2019solitary}. A generalized
solitary state, where several oscillators exhibit dynamics different  from that of the
synchronous cluster received  attention
in~Refs.~\cite{kapitaniak2014imperfect, hizanidis2016chimera,
rybalova2017transition, semenova2017coherenceincoherence, jaros2018solitary,
semenova2018mechanism, shepelev2018chimera, rybalova2018mechanism,
mikhaylenko2019weak, sathiyadevi2019long}.  This state appears at the border
between synchrony and asynchrony, as soon as repulsion starts to prevail over
attraction.  Our setup is an extension of the finite-size two-group Kuramoto
model treated in Ref.~\cite{maistrenko2014solitary}, where all oscillators were
identical.

We demonstrate that for small frequency mismatches between the groups and a weak
repulsion, there appears a small region, where the attractive units build a
synchronous cluster, while the repulsive oscillators exhibit quasiperiodic
partially synchronous dynamics. Slightly stronger repulsion leads to the
solitary state, which is replaced by quasiperiodic dynamics again for bigger
repulsion.  For large mismatches in the frequency the solitary state is not
observed, but only quasiperiodic dynamics.

\section{The Model}
A popular version of the standard  Kuramoto-Sakaguchi model is a system of $M$
interacting groups of identical units, described by the following equations:
\begin{equation}
  \dot{\t}^{\sigma}_j = \w_{\sigma} + \sum_{\sigma' = 1}^M
    \frac{K_{\sigma\sigma'}}{N} \sum_{k=1}^{N_{\sigma'}} \sin(\t^{\sigma'}_k -
    \t^{\sigma}_j + \alpha_{\sigma\sigma'}) \; ,
\end{equation}
where $\t^{\sigma}_j$ is the phase of the $i$th oscillator in the group $\sigma$
and $\sigma = 1,\ldots,M$.  Here $\w_{\sigma}$ and $N_{\sigma}$  are  the
natural frequency and the number of oscillators in the group $\sigma$,
$N=\sum_\sigma N_\sigma$, and $K_{\sigma\sigma'}$ and $\alpha_{\sigma\sigma'}$
are respectively the strength of the coupling and the phase shift characterizing
interaction between groups $\sigma$ and $\sigma'$.

In the following we analyze a two-group Kuramoto-Sakaguchi model wherein the
coupling coefficients and the phase shift parameters depend on the acting group
only, i.e. $K_{\sigma\sigma'} = K_{\sigma'}$ and $\alpha_{\sigma\sigma'} =
\alpha_{\sigma'}$.  We concentrate on a particular case, motivated by
neuroscience applications,  when the coupling within the first group is
attractive while in the second group it is repulsive.  We denote phases of the
units in these groups by $\vp$ and $\p$, respectively.  By re-scaling the time
and performing a transformation to a reference frame co-rotating with the
frequency of the attractive group, we write the model as
\begin{equation}
  \begin{aligned}
    \dot\vp_j &= \frac{1}{N}\sum_{k=1}^{N_a}\sin(\vp_k-\vp_j+\alpha_a)
      -\frac{1+\e}{N}\sum_{k=1}^{N_r}\sin(\p_k-\vp_j+\alpha_r) \; , \\
    \dot\p_j &=\w+\frac{1}{N}\sum_{k=1}^{N_a}\sin(\vp_k-\p_j+\alpha_a)
      -\frac{1+\e}{N}\sum_{k=1}^{N_r}\sin(\p_k-\p_j+\alpha_r) \; ,
  \end{aligned}\label{eq:orig_system}
\end{equation}
where subscripts $a$ and $r$ stand for ``attractive'' and ``repulsive'',
respectively. Quantification of coupling has been reduced to a single parameter
$K_r/K_a = -(1 + \e)$, with $\e$ being the excess of repulsive coupling. An $\e
< -1$ indicates that interaction within both groups is attractive and,
trivially, the whole system synchronizes.  For $\e =-1$ the second group is
uncoupled and in the range $-1 < \e < 0$ the repulsive coupling is weaker than
the attractive coupling. For $\e = 0$ their magnitudes are identical and for $\e
> 0$ the repulsive coupling dominates.

Introducing the Kuramoto mean fields for both groups, $Z_a = \r_a e^{i\T_a} =
1/N_a \sum e^{i\vp_j}$, $Z_r= \r_r e^{i\T_r} = 1/N_r \sum e^{i\p_j}$, and the
common forcing
\begin{equation}
  H = he^{i\Phi} =
    \frac{N_a}{N}e^{i\alpha_a}Z_a -\frac{N_r}{N}(1+\e)e^{i\alpha_r}Z_r \; ,
    \label{eq:common_force}
\end{equation}
we re-write the model in a compact form as
\begin{align}
  \dot\vp_j & = \im\left [ He^{-i\vp_j}\right ] = h\sin(\Vp-\vp_j) \;
    ,\label{eq:model_compact1} \\
  \dot\p_j & = \w+ \mbox{Im}\left [He^{-i\p_j}\right ] = \w+h\sin(\Vp-\p_j) \; .
    \label{eq:model_compact2}
\end{align}
For the further analysis we restrict ourselves to the case of
equally sized groups $N_r = N_a = N/2$ and $\alpha_a = \alpha_r = 0$.
Equation (\ref{eq:common_force}) then reduces to
\begin{equation}
  H = he^{i\Phi} =
    \frac{1}{2}\left [Z_a -(1+\e)Z_r\right ]  \; .
    \label{eq:common_force_r}
\end{equation}

We notice that according to the Watanabe-Strogatz (WS) theory
\cite{watanabe1993integrability, watanabe1994constants} the dynamical
description of $n>3$ identical oscillators subject to a common force can be
reduced to equations for three global variables and $n-3$ constants of motion.
Thus, for $N_{a,r}>3$ and $\w \neq 0$ the model
(\ref{eq:model_compact1},\ref{eq:model_compact2}) is in fact 6-dimensional and
can be described by two coupled systems of WS equations, see
Ref.~\cite{pikovsky2008partially}. For $\w=0$ all oscillators become identical
and the whole ensemble can be described by three WS equations.

\section{Synchronous state}

\begin{figure}
  \includegraphics{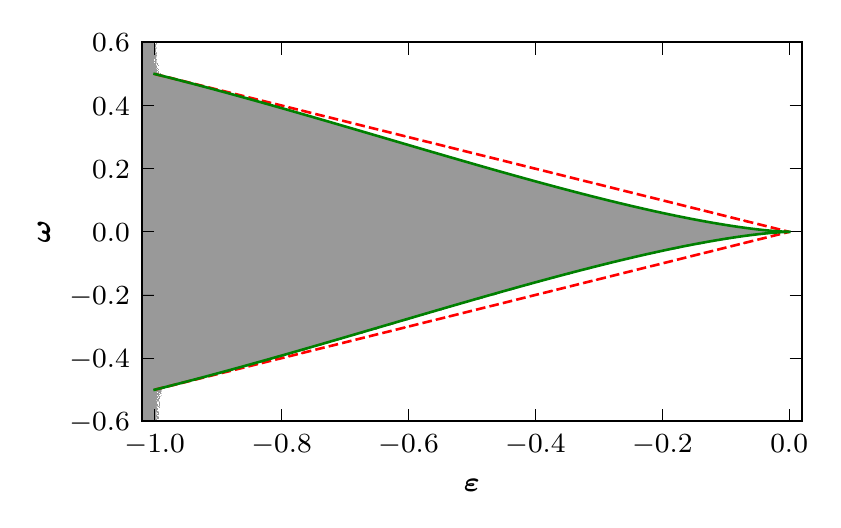}
  \caption{Full synchrony in system
  (\ref{eq:model_compact1},\ref{eq:model_compact2}) is a two-cluster state. The
  region of full synchrony, as obtained numerically, is shaded with gray, while
  all other states are shown with white.  The dashed red and the solid green
  lines show the analytical results for boundary of existence and of stability
  of the two-cluster state, respectively, see
  Eqs.~(\ref{eq:analytical_full_synchrony},\ref{eq:stability_full_synchrony}).
  }\label{fig:full_synchrony}
\end{figure}

First we analyze conditions of existence and stability of a synchronous state,
where $\vp_j = \vp$ and $\p_j = \p$ for all $j$ and observed frequencies are
$\dot{\vp} = \dot{\p} = \nu$. Notice that generally $\vp \neq \p$, i.e.\
synchrony in this setup shall be understood as existence of a two-cluster state.
Notice also that for  $\e < -1$ both groups are attractive and
synchronize regardless of $\w$, therefore we are interested in the interval $\e
> -1$. Let  $\vp = \nu t$, $\p =  \nu t + \p_0$, and $\Vp = \nu t + \Vp_0$. Then
real and imaginary parts of Eq.~(\ref{eq:common_force_r}) provide
\begin{equation}
  \begin{aligned}
    h \cos\Vp_0 & = \frac{1}{2} - \frac{1+\e}{2} \cos\p_0 \; , \\
    h \sin\Vp_0 & = - \frac{1+\e}{2} \sin\p_0 \; .
  \end{aligned}
\label{eq:fullsyn2}
\end{equation}

\paragraph{Condition of existence.}
Subtracting Eq.~(\ref{eq:model_compact1}) from Eq.~(\ref{eq:model_compact2}) and
using $\dot\p_0=0$ we find that
\begin{equation}
  \w= h[\sin\Vp_0 - \sin(\Vp_0 - \p_0)]\;. \label{eq:omega}
\end{equation}
Writing the second term as $\sin\Vp_0\cos\p_0-\cos\Vp_0\sin\p_0$ and excluding
$\sin\Vp_0$ and $\cos\Vp_0$ using Eqs.~(\ref{eq:fullsyn2}) we obtain
\begin{equation}
  \sin\p_0 = -\frac{2\w}{\e} \;. \label{eq:analytical_full_synchrony}
\end{equation}
It follows, that synchrony does not exist for $\e=0$, when attraction and
repulsion are balanced. For $\e>0$ the repulsion becomes stronger than
attraction and therefore the synchronous two-cluster state cannot be expected
either. This consideration yields the border of the synchronous domain for
$\e<0$:
\begin{equation}
  \vert \w \vert \leq - \e /2 \;. \label{eq:full_synchrony_w}
\end{equation}
In order to find the observed frequency $\nu$ we expand
(\ref{eq:model_compact2}) and insert (\ref{eq:fullsyn2}).  Together with
(\ref{eq:analytical_full_synchrony}) this yields
\begin{equation}
  \nu = \frac{1+\e}{\e} \w \; . \label{eq:synchronous_nu}
\end{equation}
Notice that the ratio $(1+\e)/\e$ is negative in the region of existence, so
that two synchronous clusters rotate in the direction, opposite to the one
determined by $\w$. (We remind that we consider the motion in a frame,
co-rotating with the natural frequency of the attractive group.)

\paragraph{Condition of stability.}
The next step is to determine stability of the two-cluster configuration. For
this purpose we first consider the linear stability of the repulsive cluster
with respect to a symmetric perturbation~\cite{yeldesbay2014chimeralike}. It
means that phases of two perturbed oscillators become $\p_{\pm} = \nu t+\p_0 \pm
\alpha$, where $\alpha\ll 1$. This assures that the mean field $Z_r$ remains
unchanged in the first-order approximation in $\alpha$. The perturbed
oscillators then evolve according to
\begin{equation}
  \dot{\p}_{\pm} = \w + h\sin(\Vp_0 - \p_0 \mp \alpha) \; .\label{eq:linear_stability}
\end{equation}
In the first order in $\alpha$ we find
\begin{equation}
  \dot{\alpha} = - \alpha h \cos(\Vp_0 - \p_0)  \; .\label{eq:stability_alpha}
\end{equation}
Thus, the cluster is stable for $h\cos(\Vp_0 - \p_0) > 0$.
With the help of Eqs.~(\ref{eq:fullsyn2}) this condition can be re-written as
\begin{equation}
 \cos\p_0 - (1+\e) >0\;.
\end{equation}
Hence, the border of stability is determined by the condition $\cos\p_0=1+\e$.
Now, using Eq.~(\ref{eq:analytical_full_synchrony}), we exclude $\p_0$ and
obtain the stability boundary as
\begin{equation}
  \w = \pm \sqrt{-\frac{\e^3}{2} - \frac{\e^4}{4}} \; .
  \label{eq:stability_full_synchrony}
\end{equation}
Using the same approach for the attractive group we find the condition for the
stability to be
\begin{equation}
  \cos\p_0 < \frac{1}{1+\e} \; .\label{eq:condition_attractive}
\end{equation}
In the domain where the synchronous state exists we have $\e<0$ and the
latter condition is fulfilled.

Next, we have to consider the stability of the two-cluster configuration with
respect to a shift of one of the clusters.  For this purpose we re-write
Eqs.~(\ref{eq:model_compact1},\ref{eq:model_compact2}) for the special case of
$\vp_j = \vp$ and $\p_j = \p$. Using Eq.~(\ref{eq:common_force}) we obtain
\begin{align}
  \dot{\vp} & = -\frac{1+\e}{2}\sin(\p - \vp) \label{eq:compact_full_1} \; , \\
  \dot{\p} & = \w - \frac{1}{2}\sin(\p - \vp) \label{eq:compact_full_2} \; ,
\end{align}
which yields the Adler equation~\cite{adler1946study} for the distance between
the clusters $\delta = \p - \vp$:
\begin{equation}
  \dot{\delta} = \w + \frac{\e}{2} \sin \delta \; . \label{eq:full_}
\end{equation}
This equation has a stable fixed point for $\vert \w \vert < -\frac{\e}{2}$,
i.e.\ in the whole domain of existence of the two-cluster solution.

The final conclusion is that the stability of the synchronous two-cluster state
is given by Eq.~(\ref{eq:stability_full_synchrony}).  This result fits very well
the numerical results shown in Fig.~\ref{fig:full_synchrony}.  As one can
see, the stable domain is smaller than the region where full synchrony exists.

\section{Nontrivial States Beyond the two-Cluster synchrony }

\subsection{Solitary State}

\begin{figure}
  \includegraphics{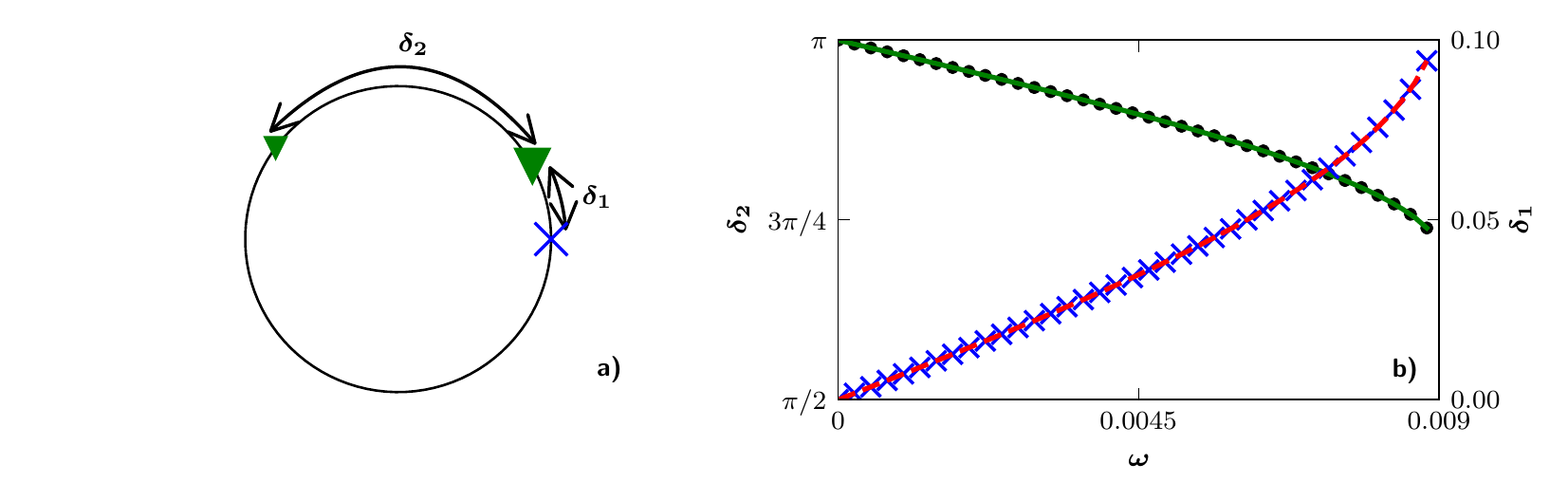}
  \caption{a) Schematic illustration of the solitary state.  Here the big  and
  small green triangles denote the cluster of $N_r-1$ repulsive units and
  solitary repulsive oscillator, respectively.  The cluster of attractive units
  is shown by the blue cross.  Panel b) shows phase differences  $\delta_{1,2}$
  in the solitary state for a particular case $N_r = N_a = 5$ and $\e=0.212$
  (this value corresponds to the largest range of  $\w$ for which  the solitary
  state exists, cf. Fig.~\ref{fig:solitary}).  Here  black circles and blue
  crosses show the results of direct numerical simulation for $\delta_2$ and
  $\delta_1$, respectively, while the solid green and the dashed red lines are
  the theoretical results obtained with the help of
  Eqs.~(\ref{eq:analytical_solitary},\ref{eq:solitary_phi}).  The boundary of
  the solitary state is at $\w \approx 0.009$.
  }\label{fig:average_frequency_solitary}
\end{figure}

The next solution we observe is the three-cluster state.  As has been shown in
Ref.~\cite{maistrenko2014solitary}, the system
(\ref{eq:model_compact1},\ref{eq:model_compact2}) with $\w=0$, exhibits, beyond
the fully synchronous one-cluster solution, a peculiar solitary state, where a
cluster of $N_a$ attractive and $N_r - 1$ repulsive oscillators coexists with a
phase-shifted solitary oscillator. This state is not of full measure, so that
not every initial condition leads to it. The range of the coupling values, where
this solution exists shrinks as $1/N$ for $N \to \infty$. This makes the
solitary state reliably observable only for small system sizes.  The picture we
observe for $\w \ne 0$ is slightly different.  Though the loss of synchrony here
also occurs via appearance of a solitary unit, now one finds a three-cluster
state: a cluster of $N_a$ attractive oscillators, a cluster of $N_r-1$ repulsive
oscillators, and a solitary repulsive unit.  The phase shifts between clusters
are constant, so that the whole configuration rotates with the same constant
observed frequency $\nu$.  An illustration of this can be found in
Fig.~\ref{fig:average_frequency_solitary}.

\begin{figure}
  \centering
  \includegraphics{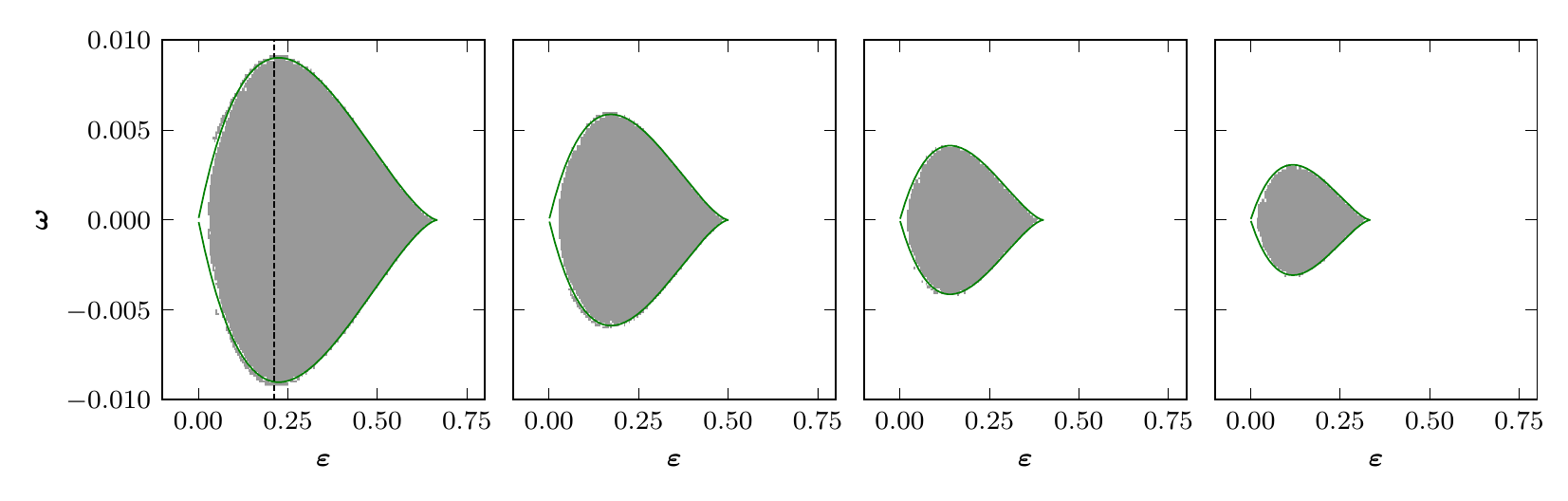}
   \caption{The solitary state is a state with three clusters of size $N_a$,
   $N_r-1$, and 1, respectively. Parameters where such a state was observed
   numerically are shaded gray, while all others are shaded white. The solid
   green line gives the analytically derived boundary, see
   Eq.~(\ref{eq:analytical_solitary}). The dashed black line in the left panel
   marks $\e=0.212$; this value approximately corresponds to the  largest
   interval of $\omega$ where the solitary state exists.  The  panels from left
   to right show the results for $N_a=N_r=5,6,7$ and $8$.}\label{fig:solitary}
\end{figure}

\paragraph{Condition of existence.}
For a description of this state we write  $\vp = \nu t$, $\p_{1,\ldots ,N_r - 1}
= \nu t + \delta_1$, $\p_{N_r} = \nu t + \delta_1 + \delta_2$, and $\Vp = \nu t
+ \Vp_0$. This yields the equations
\begin{align}
  h e^{i\Vp_0} & = \frac{1}{2} - \frac{1+\e}{2N_r} \left[(N_r-1) e^{i \delta_1} + e^{i(\delta_1 + \delta_2)} \right] \; ,
  \label{eq:solitary_forcing} \\
  \nu & = h \im[e^{i\Vp_0}] \; , \label{eq:solitary_frequency}\\
  \nu & = \w + h \im[e^{i(\Vp_0 - \delta_1)}] \; , \label{eq:solitary_alpha}\\
  \nu & = \w + h \im[e^{i(\Vp_0 - \delta_1 - \delta_2)}] \; .
\end{align}
From the last two equations it follows that  $\im[e^{i(\Vp_0 - \delta_1)}] =
\im[e^{i(\Vp_0 - \delta_1 - \delta_2)}]$ and $\re[e^{i(\Vp_0 - \delta_1)}] =
-\re[e^{i(\Vp_0 - \delta_1 - \delta_2)}]$.  This yields $2 \delta_1 + \delta_2 =
2\Vp_0 - \pi$. Multiplying (\ref{eq:solitary_forcing}) with $e^{-i\Vp_0}$ and
taking the imaginary part we find, by replacing $h$ with
Eq.~(\ref{eq:solitary_frequency}),
\begin{equation}
  \im[e^{i(\Vp_0-\delta_1)}] = \frac{1}{1+\e}\im(e^{i\Vp_0}) \; . \label{eq:solitary_phases}
\end{equation}
By applying this relation to
Eqs.~(\ref{eq:solitary_frequency},\ref{eq:solitary_alpha}) we find that observed
frequency $\nu$ is described by the same  Eq.~(\ref{eq:synchronous_nu}) as in
the synchronous state.  However, while in the case of full synchrony $(1+\e)/\e$
was negative, here it is positive.  Next,  multiplying
Eq.~(\ref{eq:solitary_forcing}) by $e^{-i\Vp_0}$ and taking this time the real
part we  obtain, after replacing $\re(e^{i\Vp_0}) = \sqrt{1 -
{\im(e^{i\Vp_0})}^2}$:
\begin{equation}
  h = \frac{1}{2}\sqrt{1-{\im(e^{i\Vp_0})}^2} - \frac{N_r-2}{2N_r}\sqrt{{(1+\e)}^2 - {\im(e^{i\Vp_0})}^2} \; .
\end{equation}
Finally, replacing $h$ with the help of
Eqs.~(\ref{eq:solitary_frequency},\ref{eq:synchronous_nu}) and introducing $x =
\im(e^{i \Vp_0})$, we obtain
\begin{equation}
  0 = x \sqrt{1 - x^2} - \frac{N_r-2}{N_r} x \sqrt{{(1+\e)}^2 - x^2} - 2 \frac{1+\e}{\e} \w \;.\label{eq:analytical_solitary}
\end{equation}
To find the parameter domain of existence of the solitary state we need to find
the range of $\w$ so that Eq.~(\ref{eq:analytical_solitary}) can be fulfilled for
a given $\e$.  First of all notice that Eq.~(\ref{eq:analytical_solitary}) is
invariant with respect to the transformation $x \to -x$ and $\w \to -\w$.  The
branch for $\w > 0$ is given by the solution for $x \in (0,1]$ and the other one
can be inferred by using the transformation $\w \to -\w$. Consider the function
$f$ consisting of the first two terms on the right hand side of
Eq.~(\ref{eq:analytical_solitary}):
\begin{equation}
  f(x, N_r, \e) = x \sqrt{1 - x^2} - \frac{N_r-2}{N_r} x \sqrt{{(1+\e)}^2 - x^2} \; . \label{eq:solitary_root}
\end{equation}
The border of the solitary state for  $\w > 0$ can then be calculated as $\w =
\frac{\e}{2(1+\e)} f_{\max}(x, N_r, \e)$.  To find the maximum of $f$ we write
$\partial{f}/\partial{x}=0$, which yields
\begin{equation}
  (1-2x^2)N_r\sqrt{{(1+\e)}^2 - x^2} = [{(1+\e)}^2 - 2x^2](N_r-2)\sqrt{1-x^2} \; . \label{eq:solitary_derivative}
\end{equation}
Squaring Eq.~(\ref{eq:solitary_derivative}) and ordering it by powers of $x$ we
get a cubic equation for $x^2$.  The expression for the roots is too long to be
shown here, but the calculated maximal $\w$ for the solitary state fits the
numerical results nicely, as shown  in Fig.~\ref{fig:solitary}.

\paragraph{Phase shifts in the solitary state.}
To determine the phase shifts $\delta_1$ and $\delta_2$, we first
rewrite Eq. (\ref{eq:orig_system}) in terms of $\delta_1$ and $\delta_2$:
\begin{equation}
  \dot{\delta}_2 = \frac{1}{2} [\sin\delta_2 ((1+\e) - \cos\delta_1) - \cos\delta_2\sin\delta_1 + \sin\delta_1] \; .
\end{equation}
Next, similarly to the case of $\w = 0$ studied in
Ref.~\cite{maistrenko2014solitary}, we write it as
\begin{equation}
  \dot{\delta}_2 = A [\sin(\delta_2 - \delta_2^*) + \sin\delta_2^*] \; ,\label{eq:solitary_delta_2}
\end{equation}
where $\tan\delta_2^* = \sin\delta_1/((1+\e) - \cos\delta_1)$ and $A =
\sin\delta_1/(2\sin\delta_2^*)$. A stable state has the solution $\delta_2 = 0$
or $\delta_2 = 2 \delta_2^* + \pi$. The first solution corresponds to the
2-cluster state and the second solution to the solitary state. As shown earlier
the phase shifts in the solitary state are related via $2 \Vp_0 - \pi =
2\delta_1 + \delta_2$. This can also be expressed as $\Vp_0 = \delta_1 +
\delta_2^*$.  Equation  (\ref{eq:solitary_phases}) then allows one to write the
relation between $\delta_2^*$ and $\Vp_0$ as
\begin{equation}
  \sin\delta_2^* = \frac{1}{1+\e}\sin\Vp_0 \; ,\label{eq:solitary_phi}
\end{equation}
and consequently allows for the calculation of $\delta_2$ and $\delta_1$ from
$\Vp_0$. $\Vp_0$ can be calculated numerically from
Eq.~(\ref{eq:analytical_solitary}) and the resulting phase shifts coincide with
the numerical results in Fig.~\ref{fig:average_frequency_solitary}.

\paragraph{Stability.}
An analytical linear stability analysis shows that the value of $\delta_2$ is
stable in the region of existence. Finding the stability for $\delta_1$ is not
as simple and can only be done numerically. Still we find it to be stable in the
whole region of existence for $N_a = N_r = 5$. The stability analysis can be
found in Appendix~\ref{sec:appendix_stability_solitary}.

\paragraph{Case $\w=0$ vs.\ case $\w\neq 0$.}
Our numerical results indicate that for $\w \neq 0$ in the parameter range where
the solitary state exists, it is the only attractor. This is an essential
difference with the previously studied case $\w = 0$, see
Ref.~\cite{maistrenko2014solitary}, where the solitary state has not full
measure. Indeed, for $\w=0$ the system
(\ref{eq:model_compact1},\ref{eq:model_compact2},\ref{eq:common_force_r}) admits
splay state solutions $h=0$ with $\T_a=\T_r=\Vp$ and
\begin{equation}
  \r_r = \r_a/(1+\e) \; . \label{eq:w0_max}
\end{equation}
For $\w\ne0$ the state $h=0$ is not a solution and numerical studies indicate
that the completely asynchronous case $\r_a=\r_r=h=0$ is unstable. Thus, the
solitary state remains the only attractor.

\paragraph{Absence of other clustered states.} According to the WS
theory~\cite{watanabe1993integrability, watanabe1994constants,
pikovsky2011dynamics}, the repulsive group can be fully described by two global
angle variables $\Psi$ and $\Gamma$, global variable $0 \leq\kappa \leq 1$, and
$N_r$ constants $\chi_k$, $k=1,\ldots,N_r$.  The latter  depend on initial
conditions and obey three additional constraints. The original phase variables
can be obtained from the global ones with the help of the M\"obius
transformation~\cite{marvel2009identical, pikovsky2015dynamics} as
$e^{i\psi_k}=e^{i\Gamma}(\kappa+e^{i(\chi_k-\Psi)})/(\kappa
e^{i(\chi_k-\Psi)}+1)$.  For $\kappa<1$, general initial conditions, i.e.\
different $\chi_k$, yield different $\psi_k$ (for an example of such dynamics
see the partially synchronous state described in the next Section). For
$\kappa=1$ typically all $\psi_k=\psi$, i.e.\ one observes a one-cluster state.
However, it is possible that $e^{i(\chi_k-\Psi)}=-1$ for some $k=n$ and then one phase
$\psi_n$ differs from other clustered phases, i.e.\ the solitary state is
observed~\cite{pikovskyprivate, maistrenko2014solitary}.  Other cluster states except for
full synchrony and the $(N_r-1,1)$ configuration are therefore not allowed, see
Ref.~\cite{engelbrecht2014classification} for a rigorous proof.  Certainly,
similar consideration can be applied to the attractive group, but there the
solitary state is unstable and only the trivial one-cluster state is observed.

\subsection{Self-Consistent Partial Synchronization}\label{sec:scps}

\subsubsection{Numerical analysis}
\begin{figure}
 \includegraphics{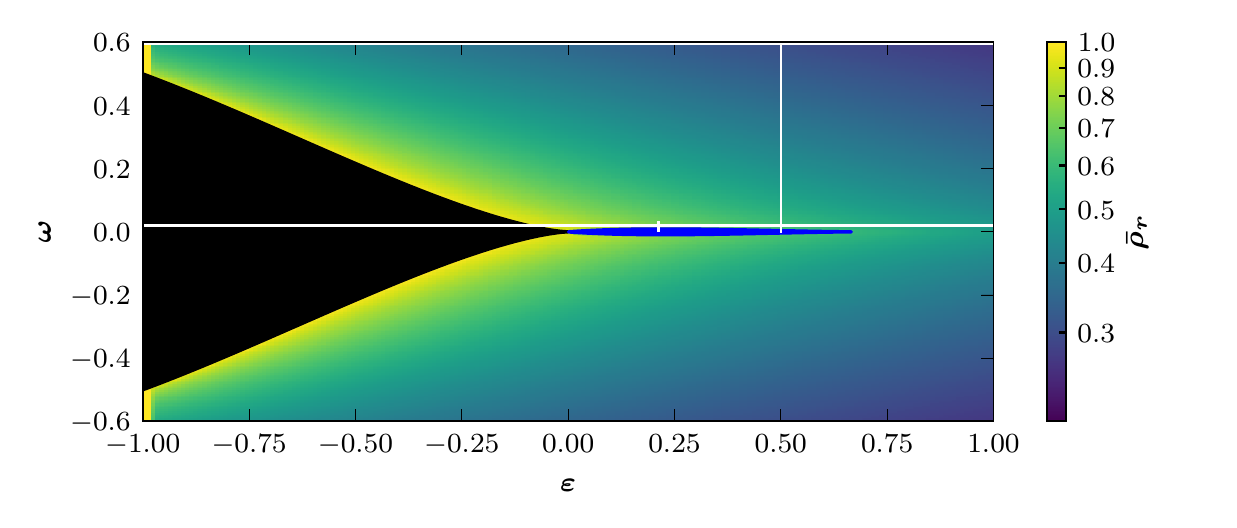}
  \caption{Overview of the states in the parameter space for $N_a=N_r=5$. The
  black region corresponds to the domain of full synchrony as determined by
  Eq.~(\ref{eq:stability_full_synchrony}). The blue color shows the domain of
  solitary states, see Eq.~(\ref{eq:analytical_solitary}).  The background
  outside of these two regions shows the time-averaged order parameter $\bar\r_r
  $ of the repulsive group for one initial condition; here it is $\bar\r_r<1$,
  so that this is the domain of partial synchrony.  White lines show  parameter
  values where the dynamics is analyzed in details, see
  Figs.~\ref{fig:frequency_oa},\ref{fig:analytical_order}.
  }\label{fig:overview}
\end{figure}

Outside of the domains of full synchrony and solitary states we find a partially
synchronized repulsive group, characterized by the order parameter $0<\r_r<1$.
As for the attractive group, we find that it remains synchronous even for such
large values of $\e$ as 10. Though the condition of its full synchrony
(\ref{eq:condition_attractive}) can be easily extended for the general case of
$\r_r \leq 1$ to $\r_r \cos(\T_r - \T_a) < (1+\e)^{-1}$, we were not able to
prove the synchrony analytically and only checked it numerically~\footnote{The
attractive group remained fully synchronized even when the units were made
non-identical by sampling the frequencies from a normal distribution with zero
mean and standard deviation of $10^{-3}$.  Hence, stability of the attractive
group is not a numerical artifact.}.  A diagram of the states, including the
domains of existence of full synchrony and of the solitary state, combined with
the presentation of the time-averaged order parameter $\bar\r_r$~\footnote{In
the following the time-averaged quantities are denoted by overlined letters.}
can be found in Fig.~\ref{fig:overview}.

\begin{figure}
  \includegraphics{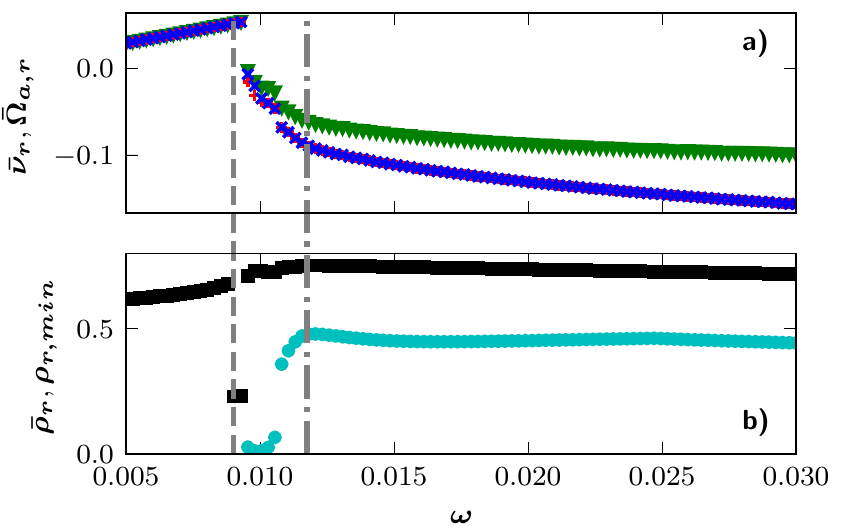}
   \caption{a) Observed frequencies of oscillators from the repulsive group,
   $\bar\nu_r$ (green triangle), and of the mean fields $\bar\Omega_r$ (blue
   crosses) and $\bar\Omega_a$ (red pluses), for $N_a=N_r = 5$ and $\e=0.212$.
   b) the average order parameter of the  repulsive group $\bar\r_r$  (black
   squares) and the minimal value of this order parameter over a long time
   interval, $\r_{r,min}=\text{min}_t[\r_r(t)]$ (cyan circles).  The analytical
   border of the solitary state is denoted by a dashed gray line; to the left of
   this line all frequencies coincide.  Close to the border of the solitary
   state, the phase is not well defined, probably due to long transients, as
   indicated by low values of $\r_{r,min}$.  This leads to the discrepancies
   between $\bar\Omega_a$ and $\bar\Omega_r$.  The right border of this domain
   is (quite arbitrary) marked by a dotted line.  To the right of this border
   $\bar\Omega_r=\bar\Omega_a$ (blue crosses and red pluses overlap).  The
   results  have been obtained taking a perturbed cluster as initial condition.
   }\label{fig:average_frequency}
\end{figure}

The observed partial synchrony can be seen as a self-organized quasiperiodic
state, SOQ (or self-consistent partial synchrony,
SCPS)~\cite{rosenblum2007self, pikovsky2009self, clusella2016minimal}. The latter
is characterized by the difference between the average frequency of the
oscillators and their mean field. Indeed, in our setup the average frequency
(observed frequency) $\bar\nu_r$ of repulsive units  is larger than the average
frequency $\bar\Omega_r $ of their mean field.  (In fact, the instantaneous
frequencies also differ nearly all the time.) Furthermore, the mismatch
$\bar\nu-\bar\Omega_r $ increases with $\w$.  Nevertheless, both sub-populations
remain synchronous on the macroscopic level, i.e.\ the average mean field
frequencies coincide, $\bar\Omega_r =\bar\Omega_a$, see
Fig.~\ref{fig:average_frequency}.  We notice that close to the border of the
solitary state these frequencies are not always well-defined, as indicated by
small values of the minimal instantaneous order parameter.  In this border
domain we observe very long transients; precise identification of the dynamical
states here requires a separate investigation.

\begin{figure}
  \centering
  \includegraphics{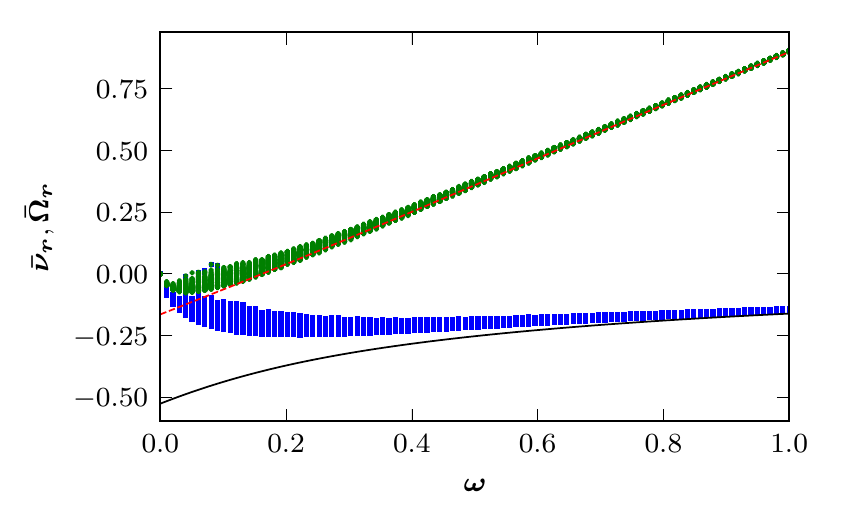}
  \caption{The observed average frequency for $N_{a,r}=5$ and $\e=0.5$, for 100
  random initial conditions per $\w$. The blue squares denote frequency $\bar\Omega_r$ of
  the mean field and the green dots denote frequency $\bar \nu_r $ of the repulsive oscillators.
  The dashed red line is the solution of Eq.~(\ref{eq:OA_freq}) and the solid
  black line is the solution for the mean field frequency, see
  Eq.~(\ref{eq:ws_3dim_tr}).}\label{fig:frequency_oa}
\end{figure}

Results for similar computations for a large range of $\omega$ are presented in
Fig.~\ref{fig:frequency_oa}. However, here the simulations were started from
many different initial conditions.  As one can see,  partially synchronous
states are characterized by a large degree of multistability: In fact, the whole
range of SCPS is multistable, as can be seen in Fig.~\ref{fig:frequency_oa} as
well as in Fig.~\ref{fig:analytical_order} below.  Different initial conditions
result in different values  of $\bar\Omega_r$ and $\bar \nu_r$~\footnote{To
obtain these quantities we have averaged the frequencies over the time interval
of 500 units, after transient of 1000 units.}.  Interestingly, the variation of
these quantities reduces with increasing $\w$.  For all these parameters the mean
fields of both populations remain synchronized; we have also checked that their
phases remain well-defined~\footnote{Even for such large values as $\w=1$ and $\e=1$ the
smallest observed order parameter over 100 different initial conditions was
0.08, with the average being 0.2.}.

Notice that transition from the solitary state to partial synchrony is
accompanied by change of the direction of rotation with respect to the considered
coordinate frame~\footnote{We remind that we use the frame, co-rotating with the
natural frequency of oscillators in  the attractive group.}.  Indeed, before the
transition all frequencies are positive, while immediately after it they are
negative, see Fig.~\ref{fig:average_frequency}.  With a further increase of the
parameter $\w$, the frequency of the repulsive units $\bar \nu_r$ becomes
positive and then tends to $\w$.  In fact, for large $\w$ or for strongly
repulsive systems, the repulsive units tend to have a uniform distribution of
phases. However, they remain perturbed by the field of the synchronous
attractive cluster, so that the uniform distribution can be reached only
asymptotically.

\begin{figure}
  \includegraphics{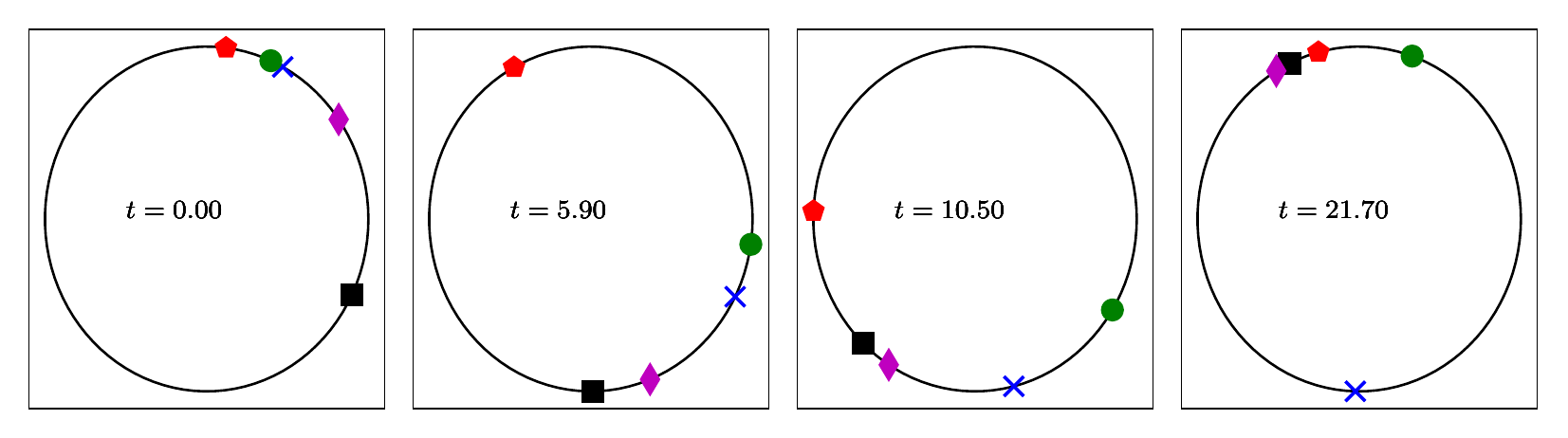}
  \caption{A specific type of partial synchrony found in the system for
  intermediate values of $\w$. The snapshots shows the repulsive oscillators
  over time, where every oscillator is marked by a different color and symbol.
  At times $t=5.9$ and $t=21.7$ a single oscillator leaves the fuzzy cluster.
  The observed system is rather small with $N_r=N_a=5$, $\e=0.2$, and
  $\w=0.1$.}\label{fig:transient_solitary}
\end{figure}

We illustrate partially synchronous dynamics of the repulsive group by several
snapshots in Fig.~\ref{fig:transient_solitary}, for an  intermediate value
$\w=0.1$.  We  see that repulsive oscillators form a group (a loose cluster),
then the first oscillator in the group accelerates, stays for some instant in
anti-phase with respect to others, so that we can speak about transient solitary
state, and then joins the group again, now becoming the last one in the group.
Then the group dissolves again, and now the oscillator that was initially the
third in the group stays for some time in anti-phase to the rest of the group,
then the group recombines, and so on.  Notice that  only every second oscillator
undergoes the transient solitary state. This dynamics seem to be independent of
the initial condition  and was observed both for even and odd $N_{a,r}$.  This
bears some resemblance to a phenomenon observed in an ensemble of attractive and
repulsive active rotators, see Ref.~\cite{zaks2016onset}.

\begin{figure}
  \includegraphics{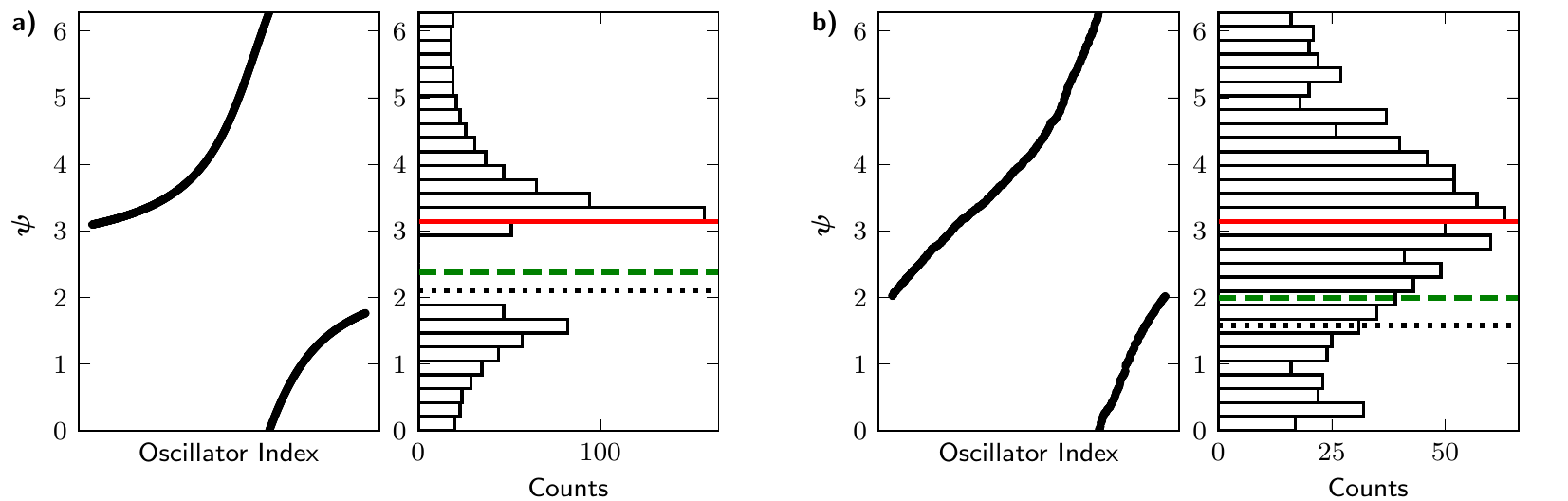}
  \caption{Phases of the repulsive group and their histograms, for two different
  initial conditions and $N_{a,r} = 1024$, $\w = 0.75$, and  $\e = 0.5$.
  Initial conditions are perturbed cluster (a) and random (b).  The solid red
  (dashed green) line is the phase of the repulsive (attractive) mean field
  $\T_r$ ($\T_a$), while the dotted black line denotes the phase of the forcing
  $\Vp$.  The distribution in a) changes its width with time,  whereas the
  distribution in b) is practically stationary.}\label{fig:multistability}
\end{figure}

To conclude the discussion of the multistability of the partially synchronous
state, we analyze a large system. In Fig.~\ref{fig:multistability} we show two
distributions of phases  $\psi$ for $N_{a,r} =1024$. These distributions have
been obtained by simulation started from different initial conditions: in one
case, illustrated in a), we use a perturbed  cluster state, while the case in b)
corresponds to random initial conditions.  The distributions differ in their
form, as well as in their dynamics.  In the first case the distribution is
bounded and bimodal; it moves with time and ``breathes'', changing its width.
Generally the phase differences between the mean fields and common force vary
in time.  In the case of random initial conditions phases spread around the unit
circle and their distribution is unimodal and nearly stationary (small time
fluctuations are probably due to finite size effect).  The differences between
the distributions also lead to slight differences in the average frequencies.
For the perturbed cluster we find $\bar \nu_r = 0.262$ and $\bar \Omega_r =
-0.196$ and for the random initial conditions we obtain $\bar \nu_r = 0.264$ and
$\bar \Omega_r = -0.181$.

\subsubsection{Theoretical analysis}
Here we provide some analytical estimates for the state of partial synchrony.
As already mentioned, for the case $\w=0$ and partial synchrony of the repulsive
units, the relation between order parameters of two groups is given by
Eq.~(\ref{eq:w0_max}).  Since the attractive group is always synchronized,
$\r_a=1$, we obtain $\r_r=1/(1+\e)$.  We expect that this expression can be used
as an estimation also for small $\w$.  We also expect that this expression
yields the upper limit for $\r_r$, since an increase in $\w$ can only lead to a
decrease in the level of synchrony.

\begin{figure}
  \includegraphics{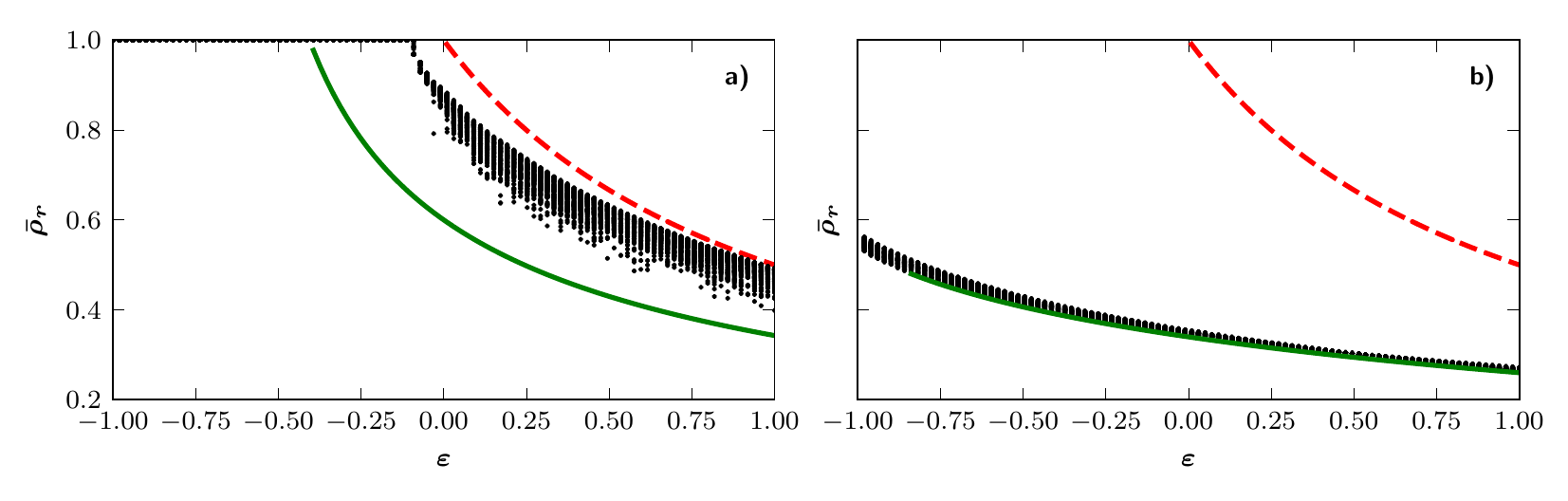}
  \caption{The average repulsive order parameter $\bar \r_r$ (black dots) for
  100 different initial conditions per $\e$ and $N_r = 5$. In a) $\w = 0.02$ and
  in b) $\w=0.6$. The dashed red line marks $1/(1+\e)$; as expected this curve
  yields a reasonable upper bound estimate for small $\w$.  The solid green line
  the solution of Eq.~(\ref{eq:WS_delta}); this estimation works better for
  large $\w$.}\label{fig:analytical_order}
\end{figure}

Next, we recall that according to the WS theory the description of the system
(\ref{eq:model_compact1},\ref{eq:model_compact2},\ref{eq:common_force_r}) can be
reduced to six equations for collective variables. (Below we use the WS
equations in the form, suggested in Ref.~\cite{pikovsky2008partially}.)
Furthermore, we restrict the consideration to the Ott-Antonsen (OA)
manifold~\cite{ott2008low, ott2009long} that corresponds to uniform distribution
of the constants of motion in the WS theory \cite{pikovsky2008partially}.  In
this case the system is further simplified, with four equations for $\r_{a,r}$
and $\Theta_{a,r}$. Moreover,  since $\r_a = 1$, we obtain a three-dimensional
system. The final equations  follow from the WS
equations~\cite{pikovsky2008partially} and read
\begin{align}
    \dot{\r_r} & = \frac{1 - \r_r^2}{4} \left[ \cos(\T_r - \T_a) - (1+\e) \r_r \right] \; , \\
    \dot{\T_r} & = \w + \frac{1 + \r_r^2}{4\r_r}  \sin(\T_a - \T_r) \; , \label{eq:ws_3dim_tr} \\
    \dot{\T_a} & = \frac{1+\e}{2} \r_r \sin(\T_r - \T_a) \; .
\end{align}
Introduction of the phase shift between the mean fields $\delta=\T_r - \T_a$ leads to
the two-dimensional system
\begin{equation}
  \begin{aligned}
    \dot{\r_r} & = \frac{1 - \r_r^2}{4} \left[ \cos\delta - (1+\e) \r_r \right] \; , \\
    \dot{\delta} & = \w - \frac{1 + [1-2(1+\e)]\r_r^2}{4\r_r}\sin\delta \; .
  \end{aligned}
  \label{eq:ws_2dim}
\end{equation}
Notice that since we are very far from the thermodynamic limit, the OA Ansatz
can be considered only as a rather crude approximation and, hence,
Eqs.~(\ref{eq:ws_2dim}) provide only some estimates.

We are interested in states, where the mean fields are locked, and therefore
$\delta$ is bounded.  We consider  a weaker condition $ \dot{\delta} = 0$ and
also neglect time variability of the order parameter, taking $\dot{\r_r} = 0$.
Applying this approximation to Eqs.~(\ref{eq:ws_2dim}) we obtain an estimation
for the average order parameter $\bar\r_r$:
\begin{equation}
  \begin{aligned}
    \cos\delta & = (1+\e)\bar\r_r \; , \\
    \sin\delta  & = \frac{4\w \bar\r_r}{1 + (1-2(1+\e))\bar\r_r^2}\; .
  \end{aligned}\label{eq:WS_delta}
\end{equation}
Eliminating $\delta$ by squaring the equations and reordering terms, we obtain a
cubic equation for $\bar\r_r^2$. The expression for the roots are too lengthy
and therefore not shown; the results for the average order parameter
$\bar\r_r$ can be seen in Fig.~\ref{fig:analytical_order}.  We see that for
large $\w$ the estimation of $\bar\r_r$ is quite good.

Given $\bar\r_r$ we find $\delta$ from Eqs.~(\ref{eq:WS_delta}). In its turn,
this yields the estimation of the average frequency of the repulsive mean field
$\bar\Omega_r$ from Eq.~(\ref{eq:ws_3dim_tr}) as
$\bar\W_r=\w+(1+\bar\r_r^2)\sin(\delta)/4\bar\r_r$.  For a known $\bar\Omega_r$
the average frequency of an oscillator can be calculated with the help of the  WS
theory~\cite{baibolatov2009periodically}. Using this we find the average
frequency $\bar \nu_r$ of the repulsive oscillators (for the derivation see
Appendix~\ref{sec:freq_OA}) to be
\begin{equation}
  \bar \nu_r = \frac{1 - \bar\r_r^2}{1 + \bar\r_r^2} \w + \frac{2 \bar\r_r^2}{1 + \bar\r_r^2} \bar\Omega_r \; .\label{eq:OA_freq}
\end{equation}
The estimated  $\bar \nu_r$  fits  the numerical results in
Fig.~\ref{fig:frequency_oa} for large $\w$ quite well; the estimate
$\bar\Omega_r$ is not as good, but also corresponds to the numerics for large
$\w$.

\section{Conclusion}
We have analyzed the interplay of attraction and repulsion in a two-group
Kuramoto model.  In the considered network each group consists of identical
elements but the groups differ in their frequencies.  We have found that if
attraction is stronger than repulsion then there exist an interval of frequency
mismatch $\w$ where the system synchronizes, in the sense that each group forms
a cluster.  The stronger the repulsion, the smaller is this interval of two
cluster synchrony.  The shift between synchronous clusters  is determined by
$\w$.  A further increase of repulsion or of $|\w|$ destroys the two-cluster
synchrony.  However, the attractive group remains synchronized while the
repulsive one undergoes a transition to quasiperiodic partial synchrony.  In
this state the order parameter of the repulsive group is between zero and one,
the mean field frequency remains locked to the frequency of the attractive
group, but individual units have a different, generally incommensurate,
frequency. For small $|\w|$ the transition from two-cluster synchrony to partial
synchrony occurs via formation of a solitary state. In this regime  there exist
two clusters (one with attractive units and one with all repulsive units but
one) and one solitary repulsive oscillator.  The borders of synchronous and
solitary regimes have been obtained analytically.  We notice that the domain of
the solitary state solutions rapidly shrinks with the increase of ensemble size,
whereas the partial synchrony persists for large ensembles as well.  For large
$|\w|$ the frequencies of the individual units and of the mean field have been
estimated with the help of the WS theory.  We believe that our results can be
useful for analysis of neuronal ensembles with excitatory and inhibitory
connections.

\begin{acknowledgements}
  This paper was developed within the scope of the IRTG 1740 / TRP 2015/50122-0,
  funded by the DFG/ FAPESP. M. R. was supported by the Russian Science
  Foundation (Grant No. 17-12-01534). The authors thank A. Pikovsky and Y.
  Maistrenko for helpful discussions.
\end{acknowledgements}

\appendix

\section{Stability of the Solitary State}\label{sec:appendix_stability_solitary}

The linear stability analysis of Eq. (\ref{eq:solitary_delta_2}) with a
perturbation of strength $\alpha$ yields
\begin{equation}
  \dot{\delta}_{2\pm} = \delta_2 \pm \alpha= A [\sin(\delta_2 - \delta_2^* \pm \alpha) + \sin\delta_2^*] \; .
\end{equation}
In the first order in $\alpha$ we find $\dot{\alpha} = - A \cos\delta_2^*
\alpha$ and thus the condition for stability is $A\cos\delta_2^* > 0$. Since $A
= \sin\delta_1/(2\sin\delta_2^*)$ this can also be written as
\begin{equation}
  \frac{\sin\delta_1}{2\tan\delta_2^*} > 0 \; .
\end{equation}
With the help of the definition of $\delta_2^*$ as $\tan\delta_2^* =
\sin\delta_1/((1+\e) - \cos\delta_1)$ the condition for stability becomes
\begin{equation}
  1+\e- \cos\delta_1 > 0 \; .
\end{equation}
Since the solitary states exists only for $\e > 0$, this condition is always
fulfilled.

To demonstrate the stability of $\delta_1$ is not that simple. The equation
for $\delta_1$ has the form
\begin{equation}
  \dot{\delta}_1 = \sin\delta_1 \left[-\frac{1}{2} + \frac{1+\e}{N} (N_r-1) + \frac{1+\e}{N} \cos\delta_2 \right] + \cos\delta_1\frac{1+\e}{N}\sin\delta_2 + \w -\frac{1+\e}{N}\sin\delta_2
\end{equation}
and cannot be reduced to a form similar to Eq.~(\ref{eq:solitary_delta_2}).  So,
we directly substitute he $\delta_{1\pm} = \delta_1 \pm \alpha$ and find
\begin{equation}
  \dot{\alpha} = \alpha \left[ \cos\delta_1 \left( -\frac{1}{2} + \frac{1+\e}{N} (N_r-1) + \frac{1+\e}{N} \cos\delta_2 \right) - \sin\delta_1 \frac{1+\e}{N}\sin\delta_2 \right] \; .
\end{equation}
We analyze this equation numerically, by computing  $\delta_1$ and $\delta_2$
with the help of Eq. (\ref{eq:solitary_phi}); this analysis shows that
$\delta_1$ is also stable.

\section{Oscillator frequency in the partially synchronous state}\label{sec:freq_OA}

The WS theory operates with  three collective variables; two of them are angles.
The first  one corresponds to the maximum of the distribution of individual
phases.  On the OA manifold this variable coincides with the phase of the mean
field.  The second angle variable (we denote it as $\chi$) determines phase
shift of individual oscillators with respect to the mean field (or, generally,
outside of the OA manifold, with respect to the first angle variable).
Correspondingly, the average frequency of units can be obtained as (see
\cite{baibolatov2009periodically} for details):
\begin{equation}
  \bar{\dot{\nu}} = \bar{\dot{\T}} - \bar{\dot\chi} \; ,
\end{equation}
where $\T,\chi$ obey the WS equations
\begin{align}
  \dot{\T} & = \w + \frac{1+\r^2}{2\r} \im [He^{-i\T}] \; , \\
  \dot{\chi} & = \frac{1-\r^2}{2\r} \im [He^{-i\T}] \; .
\end{align}
Expressing $\bar{\dot\chi}$ via $\bar{\dot\T}$ we obtain the individual frequency
\begin{equation}
 \bar{\dot\nu}  = \bar{\dot{\T}} - \frac{1-\r^2}{1+\r^2} (\bar{\dot\T} - \w)\;.
\end{equation}

\end{document}